\documentclass[preprint,showpacs,showkeys,aps,prb]{revtex4}
\usepackage[dvips]{graphicx}

\begin{document}

\title{\bf THREE LECTURES: NEMD, SPAM, and SHOCKWAVES}  

\author{
Wm. G. Hoover and Carol G. Hoover               \\
Ruby Valley Research Institute                  \\
Highway Contract 60, Box 601                    \\
Ruby Valley, Nevada 89833                       \\
}

\date{\today}

\vspace{0.1cm}

\begin{abstract}
We discuss three related subjects well suited to graduate research.  The first,
Nonequilibrium molecular dynamics or ``NEMD'', makes possible the
simulation of atomistic systems driven by external fields, subject to dynamic
constraints, and thermostated so as to yield stationary nonequilibrium
states.  The second subject, Smooth Particle Applied Mechanics or ``SPAM'', 
provides a particle method, resembling molecular dynamics, but designed to
solve
continuum problems.  The numerical work is simplified because the SPAM
particles obey ordinary, rather than partial, differential equations.
The interpolation method used with SPAM is a powerful interpretive tool
converting point particle variables to twice-differentiable field variables.
This interpolation method is vital to the study and understanding of the
third research topic we discuss, strong shockwaves in dense fluids.  Such
shockwaves exhibit stationary far-from-equilibrium states obtained with
purely reversible Hamiltonian mechanics.  The SPAM interpolation method,
applied to this molecular dynamics problem, clearly demonstrates both
the tensor character of kinetic temperature and the time-delayed response
of stress and heat flux to the strain rate and temperature gradients.
The dynamic Lyapunov instability of the shockwave problem can be analyzed in a
variety of ways, both with and without symmetry in time.
These three subjects suggest many topics suitable for graduate research in
nonlinear nonequilibrium problems.
\end{abstract}

\maketitle

\section{Thermodynamics, Statistical Mechanics, and NEMD}

\subsection{Introduction and Goals}

Most interesting systems are {\em nonequilibrium} ones, with gradients in
velocity, pressure, and temperature causing flows of mass, momentum,
and energy.  Systems with {\em large} gradients,
so that {\em nonlinear} transport is involved, are the most challenging.
The fundamental method for simulating such systems at the particle
level is nonequilibrium molecular dynamics (NEMD).\cite{b1,b2,b3}
Nonequilibrium molecular dynamics couples together Newtonian,
Hamiltonian, and Nos\'e-Hoover mechanics with thermodynamics and
continuum mechanics, with the help of Gibbs' statistical mechanics,
and Maxwell and Boltzmann's kinetic theory.  Impulsive hard-sphere
collisions or continuous interactions can both be treated.

NEMD
necessarily includes microscopic representations of the macroscopic
thermodynamic energy $E$, pressure and temperature tensors $P$ and $T$,
and heat-flux vector $Q$.  The underlying microscopic-to-macroscopic
connection is made by applying Boltzmann and Gibbs' statistical phase-space
theories, generalized to include Green and Kubo's approach to the
evaluation of transport coefficients, together with Nos\'e's approach
to introducing thermostats, ergostats, and barostats into particle
motion equations.

These temperature, energy, and pressure controls make it possible to
simulate the behavior of a wide variety of nonequilibrium flows with
generalized mechanics.  The nonequilibrium phase-space distributions
which result are typically multifractal, as is illustrated here with
a few examples taken from our website, [ http://williamhoover.info ].
These ideas are summarized in more detail in the books ``Molecular Dynamics'',
``Computational Statistical Mechanics'' and ``Time Reversibility,
Computer Simulation, and Chaos''.  The one-particle ``Galton Board''
(with impulsive forces) and the ``thermostated nonequilibrium
oscillator'' problem (with continuous forces) are simple enough for 
thorough phase-space analyses.  Macroscopic problems, like the steady
shockwave and Rayleigh-B\'enard flow, can be analyzed locally in phase
space by computing local growth rates and nonlocal Lyapunov exponents.

The main goal of all this computational work is ``understanding'',
developing simplifying
pictures of manybody systems.  The manybody systems themselves are
primarily computational entities, solutions of ordinary or partial
differential equations for model systems.  Quantum mechanics and manybody
forces are typically omitted, mostly for lack of compelling and
realistic computer algorithms.  There is an enduring gap between microscopic
simulations and realworld engineering.  The uncertainties in methods
for predicting catastrophic failures will continue to surprise us, no
matter the complexity of the computer models we use to ``understand''
systems of interest.

Number-dependence in atomistic simulations is typically small:  $1/N$
for the thermodynamic properties of periodic $N$-body systems, perhaps
$1/\sqrt{N}$ or even $1/\ln N$ in problems better treated with continuum
mechanics.  So far we have come to understand the equilibrium equation
of state, the linear transport coefficients, the Lyapunov instability of
manybody trajectories, and the irreversibility underlying the Second Law.
Improved understanding of relatively-simple hydrodynamic flows, like the
Rayleigh-B\'enard flow treated here, will follow from the special
computational techniques developed to connect different length scales.
Smooth Particle Applied Mechanics, ``SPAM'',\cite{b4,b5,b6} has proved
itself as not only a useful simulation technique for continuum systems,
but also as a powerful interpolation tool for all point-particle systems,
as is illustrated here for the shockwave problem\cite{b7,b8,b9}.

\subsection{Development of Molecular Dynamics at and Away from Equilibrium}
In the early days of expensive vacuum-tube computing the hardware and
software were largely controlled by the Federal Government and located at the
various weapons and energy laboratories at Argonne, Brookhaven, Livermore,
Los Alamos and Oak Ridge.  Fermi developed molecular dynamics at the Los
Alamos Laboratory in the summers of 1952-1953, discovering many of the
interesting nonergodic recurrence features characterizing the low-energy
behavior of one-dimensional anharmonic ``Fermi-Pasta-Ulam chains''.  The
Los Alamos Report summarizing his work was prepared a few months after his
death\cite{b10,b11,b12}.  At sufficiently low energies the anharmonic chains
showed no tendency toward equilibration while (it was discovered much
later that) at higher energies they did.
Figure 1 shows time-averaged ``mode energies'' for a six-particle
chain with two different initial conditions.  In both cases the
nearest-neighbor potential generates both linear and cubic forces:
$$
\phi(r) = (1/2)(r-1)^2 + (1/4)(r-1)^4 \ .
$$
Initially we choose the particles equally spaced and give all the energy
to Particle 1, $E = p_1^2/2$.  The left side of the Figure
corresponds to an initial momentum of 1 while the right side follows a
similarly long trajectory (100 million Runge-Kutta timesteps) starting
with the initial momentum $p_1 = 2$.  Fermi was surprised to find that
at moderate energies there was no real tendency toward equilibration
despite the anharmonic forces.  Thus the averaging techniques of
statistical mechanics can't usefully be applied to such oversimplified
systems.

\begin{figure}
\vspace{1 cm}
\includegraphics[height=8cm,width=4cm,angle=-90]{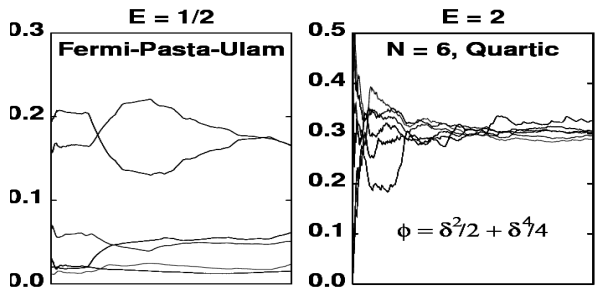}
\caption{
The time averages of the six harmonic mode energies (calculated just as
was done by Benettin\cite{b12}) are shown as functions of time for
two different initial conditions, with total energies of 0.5 and 2.0.  The
six-particle chain of unit mass particles with least-energy coordinates of
$\pm 0.5,\pm 1.5,\ {\rm and} \ \pm 2.5$ is bounded by two additional fixed
particles at $\pm 3.5$.
}
\end{figure}

Fermi also carried out some groundbreaking two-dimensional work.
He solved Newton's equations of motion,
$$
\{ \ m\ddot r = F(r) \ \} \ ,
$$
and didn't bother to describe the integration algorithm.  A likely choice
would be the time-reversible centered second-difference ``Leapfrog''
algorithm,
$$
\{ \ r_{t+dt} = 2r_t - r_{t-dt} + (F/m)_t(dt)^2 \ \} \  ,
$$
where the timestep $dt$ is a few percent of a typical vibrational
period.  The dominant error in this method is a ``phase error'', with the orbit
completing prematurely.  A harmonic oscillator
with unit mass and force constant has a vibrational period of $2\pi$.  The
second-difference Leapfrog algorithm's period is 6, rather than 6.2832, for a
relatively large timestep, $dt = 1$.  A typical set of six (repeating)
coordinate values for this timestep choice is:
$$
\{ \ +2, +1, -1, -2, -1, +1, \ \dots \ \}
$$
The Leapfrog algorithm diverges, with a period of $2\sqrt{2}$, as
$dt$ approaches $\sqrt{2}$.

Vineyard used the Leapfrog algorithm at the Brookhaven Laboratory, including
irreversible viscous quiet-boundary forces designed to minimize the effect of
surface reflections on his simulations of radiation damage\cite{b13}.  Alder and
Wainwright, at the Livermore Laboratory, studied hard disks and spheres
in parallel with Wood and Jacobsen's Monte Carlo work at the Los Alamos
laboratory, finding a melting/freezing transition for spheres\cite{b14,b15}.
  The disks
and spheres required different techniques, with impulsive instantaneous
momentum changes at discrete collision times.  All these
early simulations gave rise to a new discipline, ``molecular dynamics'',
which could be used to solve a wide variety of dynamical problems for
gases, liquids, and solids, either at, or away from, equilibrium. By
the late 1960s the results of computer simulation supported a
successful semiquantitative approach to the equilibrium thermodynamics
of simple fluids\cite{b16}.

In the 1970s Ashurst\cite{b17} (United States), Dremin\cite{b18}
(Union of Soviet Socialist Republics), Verlet\cite{b19} (France), and
Woodcock\cite{b20} (United Kingdom),  were among
those adapting molecular dynamics to the solution of nonequilibrium
problems.  Shockwaves, the subject of our third lecture, were among
the first phenomena treated in the effort to understand the challenging
problems of far-from-equilibrium many-body systems.

\subsection{Temperature Control \`a la Nos\'e}
Shuichi Nos\'e made a major advance in 1984\cite{b21,b22}, developing a
dynamics, ``Nos\'e-Hoover dynamics'', which provides sample isothermal
configurations from Gibbs' and Boltzmann's canonical distribution,
$$
f(q,p) \propto e^{-{\cal H}(q,p)/kT} \ ; \ {\cal H}(q,p) = \Phi(q) + K(p) \ .
$$
The motion equations contain one or more friction coefficients $\{ \zeta\}$
which influence the motion, forcing the longtime average of one or more of
the $p_i^2$ to be $mkT_i$:
$$
\{ \ m\ddot r_i = F_i - \zeta _ip_i \ ; \
\dot \zeta _i = [(p_i^2/mkT_i) - 1]/\tau _i^2 \ \} \ .
$$
The thermostat variable $\zeta $ can introduce or extract heat.  The
adjustable parameter $\tau$ is the characteristic time governing the
response of  the thermostat variable $\zeta$.  A useful special
case that follows from Nos\'e's work
in the limit $\tau \rightarrow 0$ is ``Gaussian'' isokinetic dynamics, a
dynamics with fixed, rather than fluctuating, kinetic energy $K(p) = K_0$,

In 1996 Dettmann showed that the Nos\'e-Hoover equations of motion follow
generally from a special Hamiltonian, without the need for the time
scaling Nos\'e used in his original work:
$$
{\cal H}_{\rm Dettmann} =
s[\Phi(q) + K(p/s) + \#kT\ln s + \#kT(p_s \tau )^2/2] \equiv 0 \ .
$$
Here the friction coefficient is $\zeta \equiv \#kT\tau^2p_s$, where
$p_s$ is the Hamiltonian momentum conjugate to $s$.  The
trick of setting the Hamiltonian equal to a special value, 0, is
essential to Dettmann's derivation\cite{b23}.

Consider the simplest interesting case, a harmonic oscillator with unit
mass, force constant, temperature, and relaxation time:
$$
{\cal H} = s[ q^2 + (p/s)^2 + \ln s^2 + p_s^2 ]/2 = 0 \rightarrow
$$
$$
\dot q = (p/s) \ ; \ \dot p = -sq \ ; \ \dot s = sp_s \ ; \
\dot p_s = -[ 0 ] + (p/s)^2 - 1 \ \rightarrow
$$
$$
\ddot q = (1/s)\dot p - (p/s)(\dot s/s) = -q -\zeta \dot q \ ; \ 
\dot p_s \equiv \dot \zeta = \dot q^2 - 1 \ .
$$
The time average of the $\dot \zeta$ equation shows that
the longtime average of $\dot q^2$ is unity.  In particular
applications $\tau $ should be chosen to maximize the efficiency of the
simulation by minimizing the necessary computer time.

Runge-Kutta integration is a particularly convenient method for solving
such sets of coupled first-order differential equations.  The fourth-order
method is
the most useful.  The time derivative is an average from four evaluations,
$\{ \ \dot y_0,\dot y_1,\dot y_2,\dot y_3 \ \} $, of the righthand
sides of all the differential equations, here collected in the form of
a single differential equation for the vector $y$:
$$
y_1 = y_0 + (dt/2)\dot y_0 \ ; \
y_2 = y_0 + (dt/2)\dot y_1 \ ; \
y_3 = y_0 + dt\dot y_2 \ ;
$$
$$
y_{dt} = y_0 + (dt/6)[\dot y_0+2\dot y_1+2\dot y_2 + \dot y_3] \ .
$$

The Runge-Kutta energy decays with time as $dt^5$
at a fixed time for a chosen timestep $dt$.  Here the vector $y$ is
$(q,p)$ so that
$$
\dot y \equiv (\dot q,\dot p) \equiv (+p,-q) \ .
$$

For small $dt$ the Runge-Kutta trajectory for a harmonic oscillator with
the exact trajectory $q = \cos(t)$ has an error
$\delta q = +dt^4t\sin(t)/120$.  The corresponding Leapfrog error is
$\delta q = -dt^2t\sin(t)/24$.
The two methods should give equally good solutions (where the two curves
in the Figure cross) when
$$
dt_{\rm LF} \simeq dt_{\rm RK}/4 = \sqrt{5/256} \simeq 0.14 \ ,
$$
corresponding to about 45 force evaluations per oscillator period\cite{b24}.

For a 14-digit-accurate trajectory calculation, with $dt_{RK} =
0.001$ and $dt_{LF} = 0.00025$, the Runge-Kutta error would be smaller than
the Leapfrog error by seven
orders of magnitude.  At the cost of additional programming complexity
choosing one of the fourth-order Gear integrators can reduce the
integration error by an additional factor of $\simeq 60$\cite{b25}.

\begin{figure}
\vspace{1 cm}
\includegraphics[height=4cm,width=4cm,angle=-90]{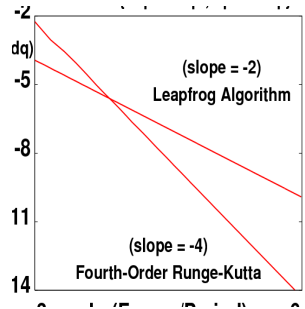}
\caption{
Comparison of the maximum error (which occurs near a time of $3\pi/2$),
in a harmonic oscillator coordinate for the Leapfrog
and Fourth-Order-Runge-Kutta integrators.  The abscissa shows the logarithm
of the number of force evaluations (which varies from about 20 to about 400)
used during a full vibrational period, $2\pi $.  The oscillator equations of
motion are $\dot q = p \ ; \ \ddot q = \dot p = -q $.
}
\end{figure}

\subsection{Connecting Microscopic Dynamics to Macroscopic Physics}
To connect the microscopic dynamics to macroscopic thermodynamics and
continuum mechanics is quite easy for a homogeneous system confined to
the volume $V$.
A numerical solution of the equations of motion for the coordinates
and momenta, 
$\{ \ q,p \ \}$, makes it possible to compute the energy $E$, the
temperature tensor $T$, the pressure tensor $P$, and the heat-flux vector
$Q$:
$$
E = \Phi(q) + K(p) = \sum_{i<j}\phi_{ij} + \sum_i p_i^2/(2m) \ ; \
$$
$$
T_{xx} = \langle p_x^2/mk \rangle = \sum_i(p_x^2/mk)_i/N \ ; \
T_{yy} = \langle p_y^2/mk \rangle \ ;
$$
$$
PV = \sum_{i<j} F_{ij}r_{ij} + \sum_i (pp/mk)_i \ ; \
$$
$$
QV = \sum_{i<j} F_{ij}\cdot p_{ij}r_{ij} + \sum_i (ep/mk)_i \ . \
$$
These expressions can be derived directly from the dynamics, by
computing the mean momentum and energy fluxes (flows per unit area and
time) in the volume $V$.  Alternatively they can be derived
by multiplying the Newtonian equations of motion by $(p/m)$ (giving
the ``Virial Theorem'') or by $e$ (giving the ``Heat Theorem'') and time
averaging\cite{b1}.  We will see that local versions of these definitions lead to
practical implementations of numerical hydrodynamics at atomistic length
and time scales.

The thermomechanical bases of these relations are statistical mechanics
and kinetic theory.  Hamilton's mechanics yields Liouville's theorem for the
time derivative of the many-body phase-space probability density
following the motion:
$$
\dot f/f = d\ln f/dt = 0 \ {\rm [Hamiltonian \ Mechanics]} \ ;
$$
Nos\'e-Hoover mechanics opens up the possibility for $f$ to change:
$$
\dot f/f = d\ln f/dt = \zeta = -\dot E/kT = \dot S_{\rm ext}/k \
{\rm [Nos\acute{e}-Hoover \ Mechanics]} \ .
$$
The primary distinction between nonequilibrium and equilibrium systems lies
in the friction coefficients $\{ \ \zeta \ \}$.  At equilibrium (ordinary
Newtonian or Hamiltonian dynamics) the average friction vanishes while in
{\em nonequilibrium} steady states
$\langle \sum k \zeta \rangle = \dot S_{\rm ext} > 0$
it is equal to the time-averaged entropy production rate.

In any stationary nonequilibrium state the sum of the friction coefficients
is necessarily positive -- a negative sum would correspond to phase-space
instability incompatible with a steady state.  An important consequence of
the positive friction is that the probability density for these states
diverges as time goes on, indicating the collapse of the probability density
onto a fractal strange attractor.  Fractals differ from Gibbs' smooth
distributions in that the  density is singular, and varies as a fractional
power of the coordinates and momenta in phase space\cite{b2,b26,b27,b28,b29}.

\subsection{Fractal Phase-Space Distributions}

The harmonic oscillator problem is not ergodic with Nos\'e-Hoover dynamics.
One way to make it so is to fix the fourth moment of the velocity distribution
as well as the second. This improvement also makes it possible to study
interesting nonequilibrium oscillator-based problems, such as the conduction
of energy from hot to cold through the oscillator motion.  Figure 3
shows the time development of (the two-dimensional projection of) such a
problem.  The isothermal oscillator, along with two friction coefficients,
$\{ \zeta , \xi \}$, fixing the second and fourth moments,
$\langle (p^2,p^4) \rangle$ has a Gaussian distribution in its
four-dimensional phase space.  A special {\em nonisothermal}
case, with a coordinate-dependent temperature leading to heat flow,
generates a 2.56-dimensional fractal in the four-dimensional
$ \{ q, p, \zeta , \xi \}$
phase space.  The dynamics governing this continuous nonequilibrium motion
is as follows:
$$
\dot q = p \ ; \ \dot p = -q - \zeta p - \xi p^3 \ ;
$$
$$
\dot \zeta = [p^2 - T] \ ; \
\dot \xi = [p^4 - 3p^2T] \ ; \
T = T(q) = 1 + \tanh(q) \ .
$$
Here time averages of the control-variable equations show that the second
and fourth moments satisfy the usual thermometric definitions:
$$
\langle p^2 \rangle = \langle T \rangle \ ; \
\langle p^4 \rangle = 3\langle p^2T \rangle \ .
$$
The phase-space distribution for this oscillator has an interesting fractal
nature\cite{b26,b27}. Figure 3 shows how the continuous trajectory
comes to give a
fractal distribution, as is typical of thermostated nonequilibrium problems.
Besides the \ae sthetic interest that this model provides, it illustrates the
possibilities for controlling moments of the velocity distribution beyond the
first and second, as well as the possibility of introducing a
coordinate-dependent temperature directly into the motion equations.

\begin{figure}
\vspace{1 cm}
\includegraphics[height=6cm,width=10cm,angle=-00]{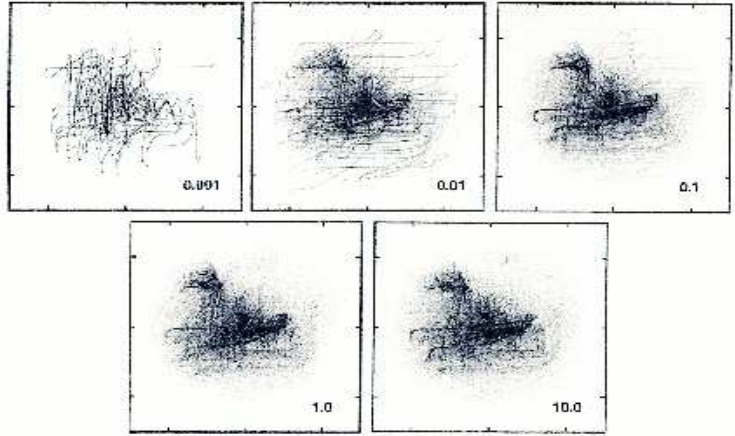}
\caption{
This $(\zeta ,\xi )$ projection of the doubly-thermostated oscillator fractal
is shown at five successive stages of temporal resolution.  The time intervals
between successive points range from 0.001, the Runge-Kutta timestep, to 10.0,
showing how a continuous trajectory can lead to a fractal object.
}
\end{figure}

\begin{figure}
\vspace{1 cm}
\includegraphics[height=4cm,width=4cm,angle=00]{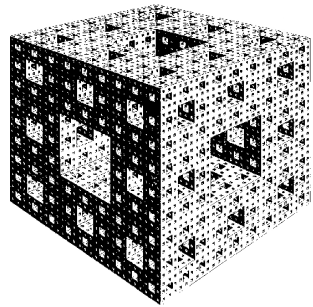}
\caption{
Sierpinski Sponge, constructed by removing 7 of the 27 equal cubes contained
in the unit cube, leaving 20 smaller cubes, and then iterating this process
{\em ad infinitum} leaving a 2.727-dimensional fractal of zero volume.
}
\end{figure}

Figure 4 shows a more typical textbook fractal, the Sierpinski sponge, in
which the probability density is concentrated on a set of dimension 2.727. In
almost all of the largest cube the density vanishes.  Unlike the multifractal
of Figure 3, the Sierpinski sponge is homogeneous, so that an $n$-fold enlarged
view of a small part of the sponge, with an overall volume $1/27^n$ of the total,
looks precisely like the entire object.

\subsection{The Galton Board}

The situation with impulsive forces is quite different.  Whenever impulsive
collisions occur the phase-space trajectory makes a jump in momentum space,
from one phase point to another.  Consider the simplest interesting case: a
single point mass, passing through a triangular lattice of hard
scatterers\cite{b2,b30,b31}.  That model generates exactly the same ergodic
dynamics as does a periodic two-hard-disk system with no center-of-mass motion:
$$
r_1 + r_2 = 0 \ ; \ v_1 + v_2 = 0 \ .
$$
By adding a constant field and an isokinetic thermostat to the field-dependent
motion, the
trajectory tends smoothly toward the field direction until a collisional jump
occurs.  Over long times (Figure 5 is based on 200,000 collisions) an
extremely interesting nonequilibrium stationary state results, with a fractal
phase-space distribution.  The example shown in the Figure has an
information dimension of 1.832.  As a consequence, the coarse-grained
entropy, $-k\langle f\ln f\rangle $, when evaluated with phase-space cells of
size $\delta $, diverges as $\delta^{-0.168}$, approaching minus infinity as
a limiting case.

The probability densities for
nonequilibrium steady states, such as the Galton Board, shown in Figure 5,
are qualitatively different to the sponge, where the probability density
is equally singular wherever it is nonzero. The Galton Board's nonequilibrium
probability density is nonzero for {\em any} configuration consistent with the
initial conditions on the dynamics. Further, the (multi)fractal dimension of
these inhomogeneous distributions varies throughout the phase space.

The concentrated nature of the nonequilibrium probability density shows first
of all that nonequilibrium states are very rare in phase space.  Finding one
by accident has probability zero.  The time reversibility of the equations
of motion additionally shows that the probability density going forward in
time contracts (onto a strange attractor), and so is necessarily stable
relative to a hypothetical reversed trajectory going backward in time, which
would expand in an unstable way.  This symmetry breaking is a microscopic
equivalent of the Second Law of Thermodynamics, a topic to which we'll
return.  It is evidently closely related to the many ``fluctuation
theorems''\cite{b32,b33} which seek to give the relative probabilities of
forward and backward nonequilibrium trajectories as calculated from
Liouville's Theorem.

\begin{figure}
\vspace{1 cm}
\includegraphics[height=4cm,width=4cm,angle=00]{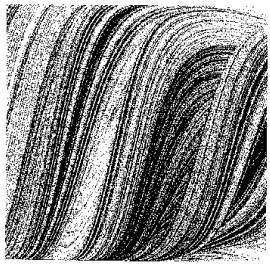}
\caption{
A series of 200,000 Galton Board collisions are plotted as separate points,
with ordinate $-1 < \sin (\beta) < 1$ and abscissa
$0 < \alpha < \pi $, where $\alpha$ is measured relative to the field
direction, as shown in Figure 6.
}
\end{figure}

\begin{figure}
\vspace{1 cm}
\includegraphics[height=4cm,width=9cm,angle=00]{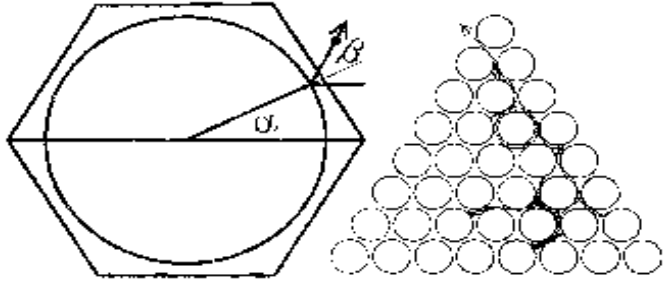}
\caption{
The Galton Board geometry is shown, defining the angles $\alpha$
and $\beta$ identifying each collision. The unit cell shown here,
extended periodically, is sufficient to describe the problem of a
moving particle in an infinite lattice of scatterers.
}
\end{figure}
\subsection{Determination of Transport Coefficients {\it via} NEMD}

With measurement comes the possibility of control. Feedback forces, based
on the results of measurement, can be used to increase or decrease a ``control
variable'' (such as the friction coefficient $\zeta$ which controls the
kinetic temperature through a ``thermostating'' force).   Equations of motion
controlling the energy, or the temperature, or the pressure, or the heat flux,
can all be developed in such a way that they are exactly consistent with
Green and Kubo's perturbation-theory of transport\cite{b2,b3}.  That theory
is a first-order perturbation theory of Gibbs' statistical mechanics.  It
expresses linear-response transport coefficients in terms of the decay of
equilibrium correlation functions.  For instance, the shear viscosity $\eta$
can be computed from the decay of the stress autocorrelation function:
$$
\eta =  (V/kT)\int_0^\infty \langle P_{xy}(0) P_{xy}(t)\rangle_{\rm eq} dt \ ,
$$
and the heat conductivity $\kappa$ can be computed from the decay of
the heat flux autocorrelation function:
$$
\kappa  = (V/kT^2)\int_0^\infty \langle Q_x(0) Q_x(t)\rangle_{\rm eq} dt \ .
$$
Nos\'e's ideas have made it possible to simulate and interpret a host
of controlled nonequilibrium situations.  A Google search for
``Nos\'e-Hoover'' in midJuly of 2010 produced over eight million
separate hits.

\begin{figure}
\vspace{1 cm}
\includegraphics[height=4cm,width=4cm,angle=-00]{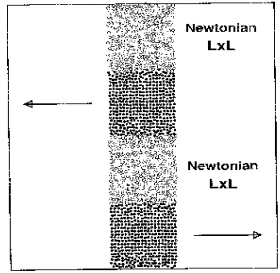}
\caption{
A Four Chamber viscous flow. Solid blocks (filled circles), move
antisymmetrically to the left and right, so as to shear the two chambers
containing Newtonian fluid (open circles).  This geometry makes it
possible to characterize the nonlinear differences among the diagonal
components of the pressure and temperature tensors.
}
\end{figure}

Figure 7 shows a relatively-simple way to obtain transport coefficients
using nonequilibrium molecular dynamics.  Ashurst\cite{b17}, in his thesis
work at the University of California, ``Dense Fluid Shear Viscosity and Thermal
Conductivity {\it via} Molecular Dynamics'', introduced two ``fluid walls'',
with different specified velocities and/or temperatures, in order
to simulate Newtonian viscosity and Fourier heat flow.  Figure 7, a
fully periodic variation of Ashurst's idea, shows two ``reservoir'' regions,
actually ``solid walls'', separating two Newtonian regions.  In both the
Newtonian regions momentum and energy fluxes react to the different velocities
and temperatures imposed in the ``wall'' reservoirs.  This four-chamber
technique produces two separate nonequilibrium profiles\cite{b34,b35,b36}.

In the Newtonian chambers, where no thermostat forces are exerted, the
velocity or temperature gradients are nearly constant, so that accurate
values of the viscosity and heat conductivity can be determined by measuring
the (necessarily constant) shear stress or the heat flux:
$$
\eta = -P_{xy}/[(dv_y/dx) + (dv_x/dy)] \ ; \ \kappa = - Q_x/(dT/dx) \ .
$$

\subsection{Nonlinear Transport}

This same ``solid-wall'' or ``four-chamber'' method has been used to study a
more complicated aspect of nonequilibrium systems, the nonlinear
contributions to the fluxes.  Because the underlying phase-space
distributions are necessarily fractal it is to be expected that
there is no analytic expansion of the transport properties analogous to the
virial (powers of the density) expansion of the equilibrium pressure.
Periodic shear flows, with the mean $x$ velocity increasing linearly with $y$,
$$
\{ \ \dot x = (p_x/m) + \dot \epsilon y \ ; \ \dot y = (p_y/m) \ \} \ ;
$$
can be generated with any one member of the family of motion equations:
$$
\{ \ \dot p_x = F_x - \dot \epsilon \alpha _x p_y - \zeta p_x \ ; \ 
\dot p_y = F_y - \dot \epsilon \alpha _y p_x - \zeta p_y \ \} \ ,
$$
so long as the sum $\alpha _x + \alpha _y$ is unity and $\zeta $ is chosen to
control the overall energy or temperature.  Careful comparisons of the two
limiting approaches,
$$
\alpha _x = 0 \ ; \ \alpha _y = 1 \ {\rm [Doll's]} \ ;
$$
$$
\alpha _x = 1 \ ; \ \alpha _y = 0 \ {\rm [s'lloD]} \ ,
$$
with corresponding boundary-driven four-chamber flows show that though both
of the algorithms satisfy the nonequilibrium energy requirement:
$$
\dot E \equiv -\dot \epsilon P_{xy}V \ ,
$$
exactly, neither of them provides the correct ``normal stress'' difference,
$P_{xx}-P_{yy}$.

This same problem highlights another interesting parallel feature of
nonequilibrium systems, the {\em tensor} nature of
temperature\cite{b7,b8,b9,b37,b38,b39,b40}.  In a
boundary-driven shearflow with the repulsive pair potential,
$$
\phi(r<1) = 100(1-r^2)^4 \ ,
$$
the temperature tensors in the Newtonian regions show the orderings
$$
\langle p^2_x\rangle > \langle p^2_z\rangle >  \langle p^2_y\rangle  \
\longleftrightarrow \
T_{xx} > T_{zz} > T_{yy} \ {\rm [Boundary \ Driven}] \ .
$$
The homogeneous periodic shear flows generated with the Doll's and
s'lloD algorithms show instead two other orderings:
$$
T_{xx} > T_{yy} > T_{zz} \ {\rm [s'lloD]} \ \ {\rm and} \ \
T_{yy} > T_{xx} > T_{zz} \ {\rm [Doll's]} \ ,
$$
so that neither the Doll's nor the s'lloD algorithm correctly accounts
for the {\em nonlinear} properties of stationary shear flows\cite{b35}.
Nonequilibrium molecular dynamics provides an extremely versatile tool
for determining nonlinear as will as linear transport.  We will come back
to tensor temperature in the third lecture, on shockwaves.  Nonlinear
transport problems can require the definition of local hydrodynamic
variables whenever the system is inhomogeneous, as it is in boundary-driven
shear and heat flows.

Thermostats, ergostats, barostats, and many other kinds of constraints and
controls simplify the treatment of complex failure problems with molecular
dynamics.  Using the Doll's and s'lloD ideas it is quite feasible to study
the stationary nonequilibrium flow of solids, ``plastic flow'', in order to
interpret nonsteady failure problems like fracture and indentation.
Nonequilibrium molecular dynamics makes it possible to remove the
irreversible heat generated by strongly nonequilibrium processes such as
the machining of metals.  The basic idea of control can be implemented from
the standpoint of Gauss' Principle, which states that the smallest possible
constraint force should be used to accomplish control\cite{b41}.  Near
equilibrium a more reliable basis is Green and Kubo's linear-response theory.
This can be used to formulate controls consistent with exact statistical
mechanics in the linear regime, just as was done in deriving the Doll's and
s'lloD approaches to simulating shear flow.

\begin{figure}
\vspace{1 cm}
\includegraphics[height=8cm,width=4cm,angle=-90]{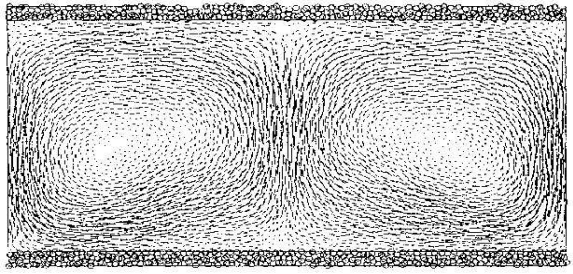}
\caption{
Rayleigh-B\'enard problem, simulated with 5000 particles.  The fluid-wall
image particles which enforce the thermal and velocity boundary conditions
are shown as circles above and below the main flow.
}
\end{figure}

A slightly more complex problem is illustrated in Figure 8.  A
nonequilibrium system with fixed mass is contained within two thermal
``fluid wall''
boundaries, hot on the bottom and cold on the top, with a gravitational
field acting downward.  If the gradients are small the fluid is stationary,
and conducts heat according to Fourier's Law.  When the Rayleigh Number,
$$
R= gL^4(d\ln T/dy)/(\nu \kappa) \ ; \ \nu \equiv \eta/\rho \ ,
$$
exceeds a critical value (which can be approximated by carrying out a linear
stability analysis of the hydrodynamic equations) two rolls, one clockwise
and the other counterclockwise provide another, faster, mode of heat transfer.
At higher values of $R$ the rolls oscillate vertically; at higher values still
the rolls are replaced by chaotic heat plumes, which move horizontally.  With
several thousand particles molecular dynamics provides solutions in good
agreement with the predictions of the Navier-Stokes-Fourier equations.

This problem\cite{b42,b43,b44} is specially interesting in that several
topologically different solutions can exist for exactly the same applied
boundary conditions.  Carol will talk more about this problem
in her exposition of Smooth Particle Applied Mechanics, ``SPAM''.  SPAM
provides a useful numerical technique for interpolating the particle
properties of nonequilibrium molecular dynamics onto convenient spatial grids.

\vspace{0.4cm}

\noindent
\section{\bf Particle-Based Continuum Mechanics \& SPAM}

\vspace{0.1cm}

\subsection{Introduction and Goals}

Smooth Particle Applied Mechanics, ``SPAM'', was invented at Cambridge,
somewhat independently, by Lucy and by Monaghan in 1977\cite{b4,b5,b6}.
The particles both men considered were astrophysical in size as their
method was designed to treat clusters of stars.  SPAM can be used on
smaller scales too.  SPAM provides a simple and versatile particle method
for solving the continuum equations numerically with a twice-differentiable
interpolation method for the various space-and-time-dependent field
variables (density, velocity, energy, ...) .  SPAM looks very much like
``Dissipative Particle Dynamics''\cite{b45}, though,
unlike DPD, it is typically fully deterministic, with no stochastic
ingredients.  Three pedagogical problems are discussed here
using SPAM: the free expansion of a compressed fluid; the collapse of a water
column under the influence of gravity; and thermally driven convection, the
Rayleigh-B\'enard problem.  Research areas well-suited to graduate research
(tensile instability, angular momentum conservation, phase separation, and
  surface tension) are also described.

SPAM provides an extremely simple particle-based solution method for solving
the conservation equations of continuum mechanics.  For a system without
external fields the basic partial differential equations we aim to solve are:
$$
\dot \rho = -\rho \nabla \cdot v \ ;
$$
$$
\rho \dot v = -\nabla \cdot P \ ;
$$
$$
\rho \dot e = -\nabla v:P -\nabla \cdot Q  \ .
$$
SPAM solves the equations by providing a particle interpretation for each
of the continuum variables occuring in these conservation laws.  The main
difficulty in applying the method involves the choice and implementation
of boundary conditions, which vary from problem to problem.

\subsection{SPAM Algorithms and the Continuity Equation}

The fluid dynamics notation here, $\{ \rho, v, e, P, Q \}$, with each of these
variables dependent on location $r$ and time $t$, is standard
but the SPAM particle interpretation of them is novel.  The density $\rho$
and momentum density $(\rho v)$ at any location $r$ are local sums of nearby
individual particle contributions,
$$
\rho (r) \equiv \sum_jm_jw(r-r_j) \ ; \ \rho (r_i) = \sum_jm_jw(r_i-r_j) \ ; \
\rho (r)v(r) \equiv \sum_jm_jv_jw(r-r_j) \ ,
$$
where particles have an extent $h$, the ``range'' of the weight function $w$,
so that only those particles within $h$ of the location $r$ contribute to the
averages there.

In the second expression (for the density {\it at} the particle location
$r_i$) the ``self'' term ($r_i = r_j$) is included so that the two definitions
coincide at the particle locations.  The weight function $w$, which describes
the spatial distribution of particle mass, or region of influence for
particle $j$, is normalized, has a smooth maximum at the origin, and a finite
range $h$, at which both $w^\prime $ and $w^{\prime \prime }$ vanish.  The
simplest polynomial filling all these needs is Lucy's\cite{b4,b5}, here
normalized for two-dimensional calculations:
$$
w_{2D}(r<h) = (5/\pi h^2)[1 - 6x^2 + 8x^3 - 3x^4] \ ; \ x \equiv r/h \ .
$$
Monaghan's weight function, shown for comparison in the Figure, uses two
different polynomials in the region where $w$ is nonzero.  The range $h$
of $w(r<h)$ is typically a scalar, chosen so that a few dozen smooth
particles contribute to the various field-point averages at a point.  As
shown in Figure 9 Lucy's function looks much like a Gaussian, but
vanishes very smoothly as $r \rightarrow h$.  By systematically introducing
the weight function into expressions for the instantaneous spatial averages of the
density, velocity, energy, pressure, and heat flux, the continuum equations
at the particle locations become ordinary differential equations much like
those of molecular dynamics. The method has the desirable characteristic
that the continuum variables have continuous first and second spatial
derivatives.

\begin{figure}
\vspace{1 cm}
\includegraphics[height=4cm,width=4cm,angle=-00]{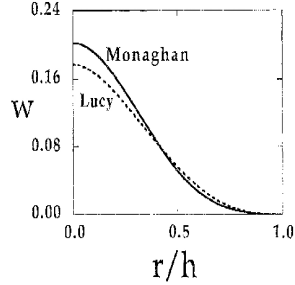}
\caption{
Lucy's and Monaghan's weight functions.  Both functions are normalized for
two space dimensions and $h=3$.  The weight function $w(r<h)$
describes the spatial influence of particles to properties in their
neighborhood, as explained in the text.
}
\end{figure}

The continuity equation (conservation of mass) is satisfied automatically.
At a fixed point $r$ in space, the time derivative of the density depends
upon the velocities of all those particles within the range $h$ of $r$:
$$
(\partial \rho /\partial t)_r \equiv \sum_j m_jv_j\cdot \nabla _jw_{rj} \equiv
-\sum_j m_jv_j \cdot \nabla _rw_{rj} \ ,
$$
where $v_j$ is the velocity of particle $j$.  On the other hand, the
divergence of the quantity $(\rho v)$ at $r$ is:
$$
\nabla _r \cdot (\rho v_r) = \nabla _r \cdot \sum_j m_jw_{rj}v_j \ ,
$$
establishing the Eulerian and Lagrangian forms of the continuity equation:
$$
(\partial \rho/\partial t)_r \equiv -\nabla _r\cdot (\rho v) \
\longleftrightarrow \ \dot \rho = -\rho \nabla \cdot v \ .
$$
These fundamental identities linking the density and velocity definitions
establish the smooth-particle method as the most ``natural''
for expressing continuous field variables in terms of particle properties.

The smooth-particle equations of motion have a form closely resembling the
equations of motion for classical molecular dynamics:
$$
\{ \ m_j\dot v_j =
-\sum_km_jm_k[(P/\rho^2)_j + (P/\rho^2)_k]\cdot \nabla _jw_{jk} \ \} \ .
$$
It is noteworthy that the field velocity at the location of particle $i$
$$
v(r=r_i) = \sum_j v_jw_{ij}/\sum_j w_{ij} =
\sum_j m_jv_jw_{ij}/\rho (r=r_i) \ ,
$$
(where the ``self'' term is again included) is usually different to the
particle velocity $v_i$, opening up the possibility
for computing velocity fluctuations {\it at} a point, as we do in the
next Section.

Notice that the simple adiabatic equation of state $P \propto \rho^2/2$ gives
exactly the same motion equations for SPAM as does molecular dynamics.  That
isomorphism pictures the weight function $w(r)$ as the equivalent of a
short-ranged purely-repulsive pair potential.  Thus the continuum
dynamics of a special two-dimensional fluid become identical to the
molecular dynamics of a dense fluid with smooth short-ranged repulsive
forces.\cite{b6} We consider this case further in applying SPAM to the free
expansion problem in the next Section.

\subsection{Free Expansion Problem}

\begin{figure}
\vspace{1 cm}
\includegraphics[height=6cm,width=4cm,angle=+90]{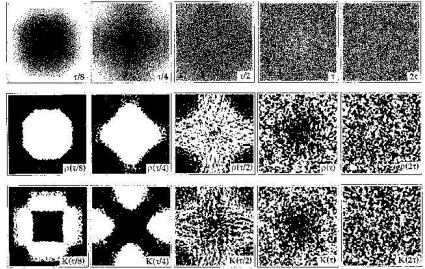}
\caption{
Contours of average density (middle row) and average temperature
(bottom row) calculated
from the instantaneous 16,384-particle snapshots (top row) taken during a
free expansion simulation.  The last picture in each row corresponds to
two sound traversal times.
}
\end{figure}

Figure 10 shows snapshots from a free expansion problem in which 16,384
particles, obeying the adiabatic equation of state $P \propto (\rho ^2/2)$,
expand to fill a space four times that of the initial compressed
gas.  This problem provides a resolution of Gibbs' Paradox (that the entropy
increases by $Nk\ln4$ while Gibbs' Liouville-based entropy,
$-k\langle \ln f \rangle $, remains unchanged)\cite{b46,b47}.  Detailed
calculations show that the missing Liouville entropy is embodied in the
kinetic-energy fluctuations.  When these fluctuations are computed in a frame
moving at the local average velocity,
$$
v(r) = \sum_jw_{rj}v_j/\sum_jw_{rj} \ , 
$$
the corresponding velocity fluctuations,
$(\langle v^2 \rangle - \langle v \rangle^2)$
are just large enough to reproduce the thermal entropy.  Most of the spatial
equilibration occurs very quickly, in just a few sound traversal times.
The contours of average density and average kinetic energy shown here
illustrate another
advantage of the SPAM averaging algorithm. The field variables are defined
{\em everywhere} in the system, so that evaluating them on a regular grid,
for plotting or analyses, is easy to do.

These local velocity fluctuations begin to be important only when the
adiabatic expansion stretches all the way across the periodic confining box so
that rightward-moving fluid collides with its leftward-moving periodic image
and {\it vice versa}.  The
thermodynamic irreversibility of that collision process, reproduced in the
thermal entropy, is just sufficient for the reversible dynamics to reflect the
irreversible entropy increase, $Nk\ln 4$.  A dense-fluid version of this
dilute-gas free expansion problem appears in Bill's lecture on shockwaves.
\begin{figure}
\vspace{1 cm}
\includegraphics[height=4cm,width=4cm,angle=-90]{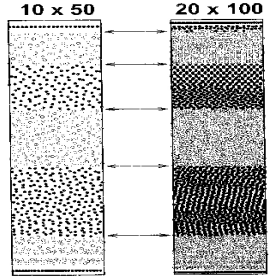}
\caption{
Equilibrated column for two system sizes.  Five density contours are indicated
by changes in plotting symbols.  The arrows corresponding to the contours
were calculated analytically from the continuum force-balance equation,
$dP/dy = - \rho g$.
}
\end{figure}

\subsection{Collapse of a Fluid Column}

\begin{figure}
\vspace{1 cm}
\includegraphics[height=4cm,width=4cm,angle=-00]{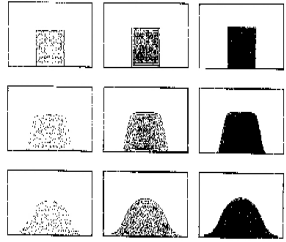}
\caption{
Water Column collapse for three system sizes  The computational time for
this two-dimensional problem varies as the three-halves power of the
number of particles used because corresponding times increase as
$\sqrt{N}$ while the number of interactions varies as $N$.
}
\end{figure}

Figure 11 shows the distribution of smooth particles in an equilibrated
periodic water column in a gravitational field\cite{b6}.  Figure 12 shows
snapshots from the subsequent collapse of the water column when the vertical
periodic boundaries are released.  Both the equilibration shown in Figure 11
and the collapse shown in Figure 12 use
the simple equation of state $P = \rho^3 - \rho^2$,
chosen to give zero pressure at unit density.  Here the gravitational
field strength has been chosen to give a maximum density of 2 at the
reflecting lower boundary.  Initially, the vertical boundaries are periodic,
preventing horizontal motion.  After a brief equilibration period, the
SPAM density profile can be compared to its analytic analog, derived by
integrating the static version of the equation of motion:
$$
dP/dy = -\rho g \ .
$$
The arrows in Figure 11, computed from the analytic static density
profile, show excellent agreement with the numerical SPAM simulation.

In smooth particle applied mechanics (SPAM) the boundary conditions are
invariably the most difficult aspect of carrying out a
simulation\cite{b6,b48}.  Here we have used a simple mirror boundary condition
at the bottom of the column and a periodic boundary at the sides, in the
vertical direction. When the vertical periodic boundary constraint is
released, rarefaction waves create a tensile region inside the falling column.
By varying the size of the smooth particles the resolution of the motion can
be enhanced, as Figure 12 shows. With ``mirror boundaries'', elaborated
in the next Section, more complicated situations can be treated.  With mirrors
there is an image particle across the boundary, opposite to each SPAM
particle, with the mirror particle's velocity and temperature both chosen to
satisfy the corresponding boundary conditions.

\begin{figure}
\vspace{1 cm}
\includegraphics[height=4cm,width=8cm,angle=-00]{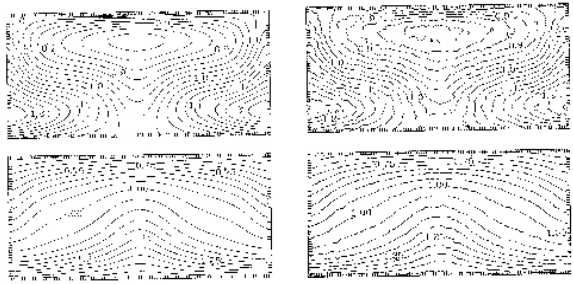}
\caption{
Instantaneous temperature (below) and density (above) contours for the
two-roll Rayleigh-B\'enard problem.  The stationary continuum solution
(left) is compared to a SPAM snapshot with 5000 particles (right).
}
\end{figure}

\subsection{Rayleigh-B\'enard Convection}

Figure 13 shows a typical snapshot for a slightly more complicated
problem,
the Rayleigh-B\'enard problem, the convective flow of a compressible fluid
in a gravitational field with the temperature specified at both the bottom
(hot) and top (cold) boundaries.  The velocities at both these boundaries must
vanish, and can be imposed by using mirror particles resembling the image
charges of electricity and magnetism.  A particularly interesting aspect of
the Rayleigh-B\'enard problem is that multiple solutions of the continuum
equations can coexist, for instance two rolls or four, with exactly the
same boundary conditions\cite{b42,b43,b44,b49}.  Such work has been used to
show that neither the entropy nor the entropy production rate allows one to
choose ``the  solution''.  Which solution is observed in practice can depend
sensitively on the initial conditions.  The Rayleigh-B\'enard problem
illustrates the need for {\em local} hydrodynamic averages describing the
anisotropies of two- and three-dimensional flows.

SPAM provides an extremely useful interpolation method for generating
twice-differentiable averages from particle data.  In the following lecture
this method will be used to analyze a dense-fluid molecular dynamics shockwave
problem, where
all of the thermomechanical variables make near-discontinuous changes linking
an incoming cold state to an outgoing hot one.  The continuous differentiable
field variables provided by SPAM make it possible to analyze the relatively
subtle nonlinear properties of such strongly nonequilibrium flow fields.

SPAM is a particularly promising field for graduate research.  In addition
to the many possible treatments of boundaries (including boundaries between
different phases), the conservation of angular momentum (when shear stresses
are present) and the tensile instability (where $w$ acts as an attractive
rather than repulsive force) and the treatment of surface tension all merit
more investigation.  For a summary of the current State of the Art see our
recent book\cite{b6}.

\vspace{0.4cm}

\noindent
\section{\bf Tensor-Temperature Shockwaves {\it via } Molecular
  Dynamics}

\subsection{Introduction and Goals}

Shockwaves are an ideal nonlinear nonequilibrium application of molecular dynamics.
The boundary
conditions are purely equilibrium and the gradients are quite large.  The
shockwave process is a practical method for obtaining high-pressure
thermodynamic data.  There are some paradoxical aspects too.  Just as in
the free expansion problem, time-reversible motion, with constant Gibbs'
entropy, describes a macroscopically irreversible process in which entropy
increases.  The increase is third-order in the compression, for weak
shocks\cite{b50}.  The shockwave problem is a compelling example of
Loschmidt's reversibility paradox.

We touch on all these aspects of the shockwave problem here. We generate and
analyze the pair of shockwaves which results from the collision of two
stress-free blocks\cite{b8,b9}.  The blocks are given initial velocities just
sufficient
to compress the two cold blocks to a hot one, at twice the initial density.
Further evolution of this atomistic system, with the initial kinetic energy of
the blocks converted to internal energy, leads to a dense-fluid version of
the free expansion problem discussed earlier for an adiabatic gas.  Here
we emphasize the dynamical reversibility and mechanical instability of this
system, show the shortcomings of the usual Navier-Stokes-Fourier description
of shockwaves, and introduce a two-temperature continuum model which
describes the strong shockwave process quite well.

\subsection{Shockwave Geometry}

There is an excellent treatment of shockwaves in Chapter IX of Landau and
Lifshitz' ``Fluid Mechanics'' text\cite{b50}.  A stationary shockwave, with steady
flow in the $x$ direction, obeys three equations for the fluxes of mass,
momentum, and energy derived from the three continuum equations
expressing the conservation of mass, momentum, and energy:
$$
\rho v = \rho _Cu_s = \rho_H(u_s-u_p) \ ;
$$
$$
P_{xx} + \rho v^2 = P_C +  \rho _Cu_s^2 = P_H +  \rho _H(u_s - u_p)^2 \ ;
$$
$$
\rho v[e + (P_{xx}/\rho) + (v^2/2)] + Q_x =
$$
$$
[e + (P_{xx}/\rho)]_C + (u_s^2/2) = 
[e + (P_{xx}/\rho)]_H + (u_s - u_p)^2/2 \ .
$$
Figure 14 illustrates the shockwave geometry in a special coordinate frame.
In this frame the shockwave is stationary.  Cold
material enters from the left at the ``shock speed'' $u_s$ and hot
material exits at the right, at speed $u_s-u_p$, where $u_p$ is the
``particle'' or ``piston'' velocity.  The terminology comes from an
alternative coordinate system, in which
motionless cold material is compressed by a piston (moving at $u_p$),
launching a shockwave (moving at $u_s$).

Eliminating the two speeds from the three conservation equations gives
the Hugoniot equation,
$$
e_H - e_C = (P_H+P_C)(V_C - V_H)/2 \ ,
$$
which relates the equilibrium pressures, volumes, and energies of the
cold and hot states.  Evidently purely equilibrium thermodynamic equation
of state information can be obtained by applying the conservations laws to
optical or electrical velocity measurements in
the highly-nonequilibrium shockwave compression process.  Ragan described the
threefold compression of a variety of materials (using an atomic bomb
explosion to provide the pressure) at pressures up to 60 Megabars, about
15 times the pressure at the center of the earth\cite{b51}.

\begin{figure}
\vspace{1 cm}
\includegraphics[height=9cm,width=3cm,angle=-90]{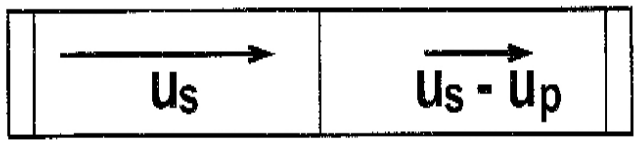}
\caption{
Stationary shockwave in the comoving frame.  Cold material enters at
the left, with velocity $+u_s$, and is decelerated by the denser hotter
material which exits at the right, with velocity $u_s-u_p$.  It is in
this coordinate frame that the fluxes given in the text are constant.
}
\end{figure}

\begin{figure}
\vspace{1 cm}
\includegraphics[height=4cm,width=6cm,angle=-00]{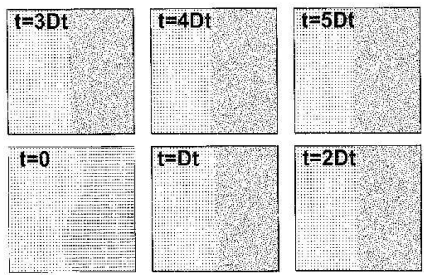}
\caption{
A series of snapshots showing the stability of a planar shockwave.  Note
that the decay of the initial sinewave profile is slightly underdamped.
Here $Dt = 2000dt$ is the time required for a shockwave to traverse the
width shown here, 2000 Runge-Kutta timesteps with $dt = 0.02/u_s \simeq
0.01$.
}
\end{figure}

\begin{figure}
\vspace{1 cm}
\includegraphics[height=4cm,width=4cm,angle=-00]{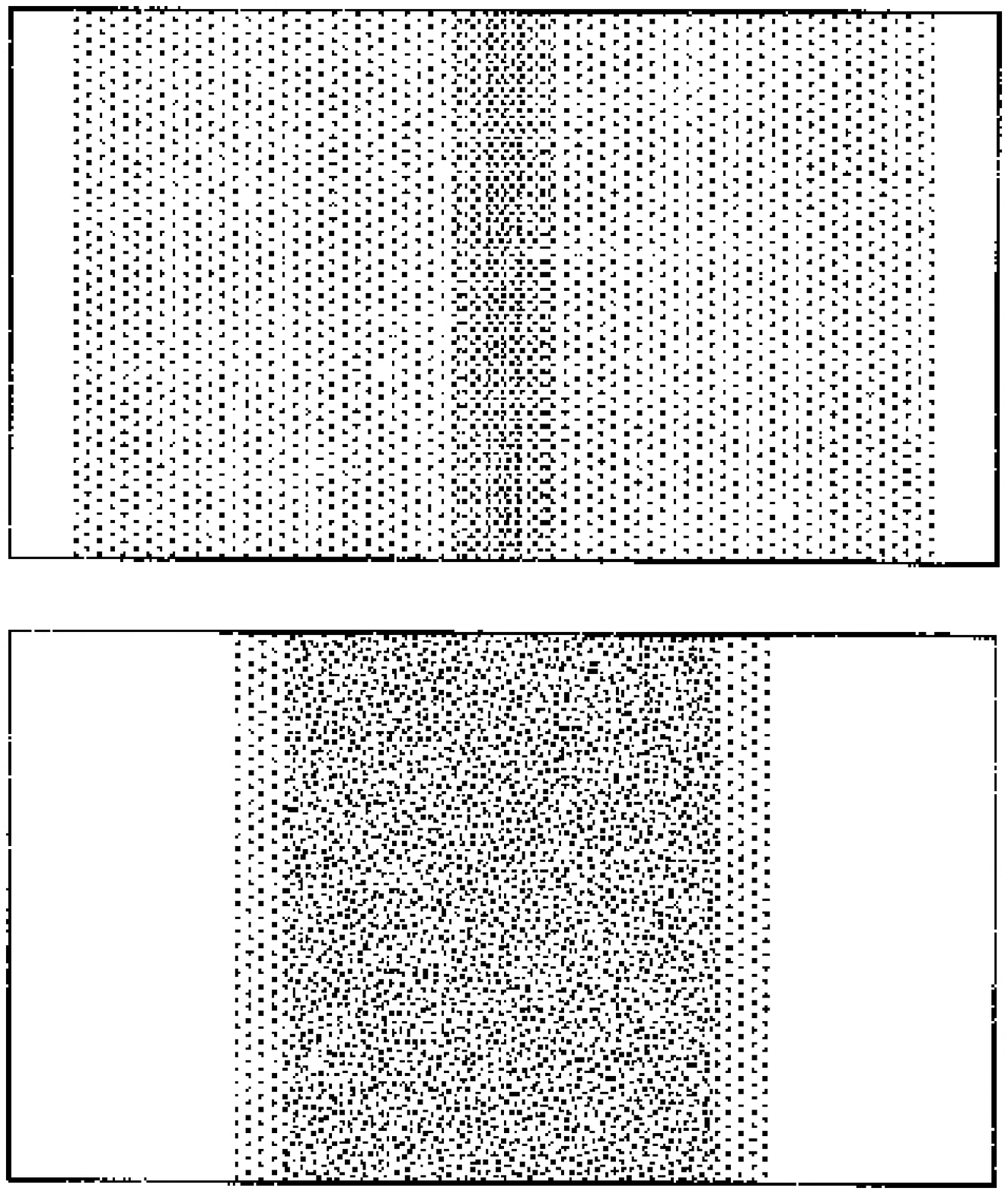}
\caption{
Snapshots near the beginning (upper) and end (lower) of an inelastic
collision between two 400-particle blocks.  The initial velocities,
$\pm 0.965$, are just sufficient for a twofold compression of the
cold material. The unit-mass particles interact with a short-ranged
potential $(10/\pi)(1-r)^3$ and have an initial density $\sqrt{4/3}$.
}
\end{figure}

Hoover carried out simulations of the shockwave compression process for a
repulsive potential, $\phi(r) = r^{-12}$, in 1967\cite{b52}, but put off
completing the
project for several years, until computer storage capacity and execution
speeds allowed for more accurate work\cite{b53}.  Comparison of Klimenko
and Dremin's computer simulations\cite{b18} using the Lennard-Jones
potential, $\phi(r) = r^{-12} - 2r^{-6}$, showed that relatively weak
shockwaves (30 kilobars for argon, 1.5-fold compression) could be
described quite well\cite{b53,b54} with the three-dimensional Navier-Stokes
equations, using Newtonian viscosity and Fourier heat conduction:
$$
P = P_{eq} - \lambda \nabla \cdot v - \eta [\nabla v + \nabla v^t] \ ;
$$
$$
\lambda_{2D} = \eta_v - \eta \ ; \ \lambda _{3D} = \eta_V - (2/3)\eta \ ;
$$
$$
Q = - \kappa \nabla T \ .
$$
Here $\lambda $ is the ``second viscosity'', defined in such a way that the
excess hydrostatic pressure due to a finite strain rate is $-\eta_V\nabla \cdot v$.
The shear viscosity $\eta$ and heat conductivity $\kappa$ were determined
independently using molecular dynamics simulations.  The small scale of the
waves\cite{b54}, just a few atomic diameters, was welcomed by high-pressure
experimentalists weary of arguing that their explosively-generated
shockwaves measured equilibrium properties.

It is necessary to verify the one-dimensional nature of the waves too.  It
turns out that shockwaves {\em do} become planar very rapidly, at nearly the
sound velocity.  The rate at which sinusoidal perturbations are damped out has
been used to determine the plastic viscosity of a variety of metals at high
pressure\cite{b55}.  Figure 15 shows the rapid approach to planarity of a
dense-fluid shockwave\cite{b9}.

Stronger shockwaves, where the bulk viscosity is more important
(400 kilobars for argon, twofold compression), showed that the Navier-Stokes
description needs improvement at higher pressures.  In particular, within
strong shockwaves temperature becomes a symmetric tensor, with
$T_{xx} >> T_{yy}$, where $x$ is again the propagation direction. In
addition, the Navier-Stokes-Fourier shockwidth, using linear transport
coefficients, is too narrow.  The tensor character of temperature in
dilute-gas shockwaves had been carefully discussed in the 1950s by
Mott-Smith\cite{b37}.

\subsection{Analysis of Instantaneous Shockwave Profiles using SPAM Averaging}

Data for systems with impulsive forces, like hard spheres, require both
time and space averaging for a comparison with traditional continuum
mechanics.  Analyses of molecular dynamics data with continuous potentials
need no time averaging, but still require a spatial smoothing operation to
convert instantaneous particle data, $\{ x,y,p_x,p_y \}_i $, including
$\{ P,Q,T,e \}_i $, to equivalent continuous continuum profiles,
$\{ \rho (r,t),v(r,t),e(r,t),P(r,t),T(r,t),Q(r,t)\} $.

The potential parts of the virial-theorem and heat-theorem expressions for
the pressure tensor P and the heat-flux vector Q,
$$
PV = \sum_{i<j} F_{ij}r_{ij} + \sum_i (pp/mk)_i \ ; \
$$
$$
QV = \sum_{i<j} F_{ij}\cdot p_{ij}r_{ij} + \sum_i (ep/mk)_i \ , \
$$
can be apportioned in at least three ``natural'' ways between pairs of
interacting particles\cite{b7,b56}.

Consider the potential energy of two particles, $\phi (|r_{12}|)$.  This
contribution to the system's energy can be split equally between the two
particle locations, $r_1$ and $r_2$, or located at the midpoint between
them, $(r_1 + r_2)/2$, or distributed uniformly\cite{b56} along the line
$r_1-r_2$ joining them.  These three possibilities can be augmented
considerably in systems
with manybody forces between particles of different masses.  It is fortunate
that for the short-ranged forces we study here the differences among the
three simpler approaches are numerically insignificant.  Once a choice has been
made, so as to define {\em particle} pressures and heat fluxes, these can in
turn be used to define the corresponding continuum field variables at any
location $r$ by using the weight-function approach of smooth particle
applied mechanics:
$$
P(r) \equiv \sum _j P_jw_{rj}/\sum _j w_{rj} \ ; \
Q(r) \equiv \sum _j Q_jw_{rj}/\sum _j w_{rj} \ .
$$

By using this approach our own simulations have characterized another
constitutive
complication of dense-fluid shockwaves -- the time delays between
[1] the maximum shear stress and the maximum strainrate and [2] the
maximum heat flux and and the maxima of the two temperature gradients
$(dT_{xx}/dx)$ and $(dT_{yy}/dx)$\cite{b8,b57}.  The study of such delays
goes back to Maxwell.  The ``Maxwell relaxation'' of a viscoelastic fluid
can be described by the model\cite{b7,b8,b57}:
$$
\sigma + \tau \dot \sigma = \eta \dot \epsilon \ .
$$
so that stress reacts to a changing strainrate after a time of order $\tau $.
Cattaneo considered the same effect for the propagation of heat.
The phenomenological delays, found in the dynamical results, are a reminder
that the irreversible nature of fluid mechanics is fundamentally different
to the purely-reversible dynamics underlying it.

The irreversible shock process is particularly interesting from the
pedagogical standpoint.  The increase in entropy stems from the conversion
of the fluid's kinetic energy density, $\rho v^2/2$ to heat.  To avoid the
need for discussing the work done by moving pistons of Figure 14, we choose
here to investigate shockwaves generated by symmetric collisions of two
stressfree blocks, periodic in the direction parallel to the shockfront.
The entropy increase is {\em large} here (a zero-temperature classical
system has an entropy of minus infinity).  Figure 16
shows two snapshots for a strong shockwave yielding twofold
compression of the initial cold zero-pressure lattice. The mechanical and
thermal variables in a strong dense-fluid shockwave are shown in Figure 17.
In order to model these results two generalizations of traditional
hydrodynamics need to be made: the tensor nature of temperature and the
delayed response of stress and heat flux both need to be treated.  A
successful approach is described next.

\begin{figure}
\vspace{1 cm}
\includegraphics[height=4cm,width=4cm,angle=-00]{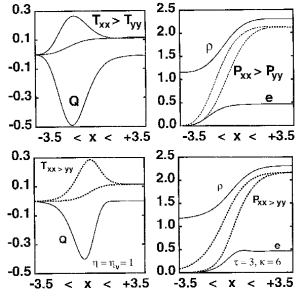}
\caption{
Shock Thermal and Mechanical Profiles from molecular dynamics are shown
at the top.  Corresponding numerical solutions of the generalized
continuum equations are shown at the bottom.  This rough comparison suggests
that the generalized equations can be fitted to particle simulations.  The
generalized equations use tensor temperature and apportion heat and work
between the two temperatures $T_{xx}$ and $T_{yy}$.  They also include
delay times for shear stress, for heat flux, and for thermal equilibration.
}
\end{figure}

\begin{figure}
\vspace{1 cm}
\includegraphics[height=7cm,width=4cm,angle=+90]{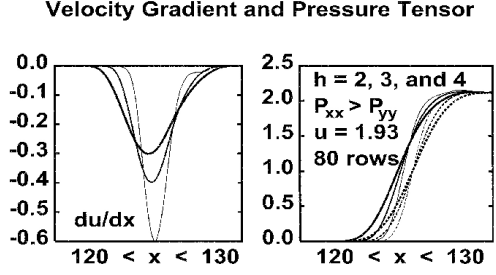}
\caption{
Shear Stress lags behind the strainrate.  The molecular dynamics gradients, using
smooth-particle interpolation, are much
more sensitive to the range of the weighting function than are the fluxes.
The results here are shown for $h=2,3,4$, with line widths corresponding to
$h$.
}
\end{figure}
\begin{figure}
\vspace{1 cm}
\includegraphics[height=7cm,width=4cm,angle=+90]{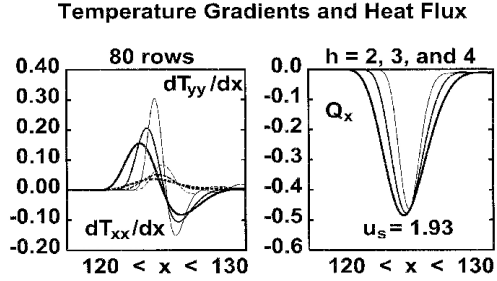}
\caption{
Heat Flux lags behind the temperature gradients.  The molecular dynamics
gradients, using smooth-particle interpolation, are much
more sensitive to the range of the weighting function than are the fluxes.
The results here are shown for $h=2,3,4$, with line widths corresponding to
$h$.
}
\end{figure}

\subsection{Macroscopic Generalizations of the  Navier-Stokes-Fourier Approach}

By generalizing continuum mechanics to include tensor temperature and the
time delays for stress and heat flux,
$$
\sigma + \tau \dot \sigma = \eta \dot \epsilon \ ; \
Q + \tau \dot Q = -\kappa \nabla T \ ,
$$
with an additional relaxation time describing the joint thermal
equilibration of $T_{xx}$ and $T_{yy}$ to a common temperature $T_H$ the
continuum and dynamical results can be made
consistent\cite{b7,b8,b9,b37,b38,b40}.
In doing this we partition the work done and the heat gained into separate
longitudinal ($x$) and transverse ($y$) parts:
$$
\rho \dot T_{xx} \propto -\alpha \nabla v:P - \beta \nabla \cdot Q 
+\rho(T_{yy} - T_{xx})/\tau \ ; \
$$
$$
\rho \dot T_{yy} \propto -(1-\alpha) \nabla v:P - (1-\beta) \nabla \cdot Q
+\rho(T_{xx} - T_{yy})/\tau \ .
$$
Solving the time-dependent continuum equations for such shockwave problems
is not difficult\cite{b9,b42}.  If all the spatial derivatives in the
continuum equations are expressed as
centered differences, with density defined in the center of a grid of
cells, and all the other variables (velocity, energy, stress, heat flux,
...) at the nodes defining the cell vertices, fourth-order Runge-Kutta
integration converges nicely to solutions of the kind shown in 
Figure 17.

\subsection{Shockwaves from Two Colliding Blocks are Nearly Reversible}

To highlight the reversibility of the irreversible shockwave process
let us consider the collision  of two blocks of two-dimensional
zero-pressure material, at a density of $\sqrt{4/3}$
(nearest-neighbor distance is unity, as is also the particle mass).
Measurement of the equation of state with ordinary Newtonian mechanics,
using the pair potential,
$$
\phi(r<1) = (10/\pi)(1-r)^3 \ ,
$$
indicates (and simulation confirms) that the two velocities $u_s$ and
$u_p$,
$$
u_s = 2u_p = 1.930 \ ,
$$
correspond to twofold compression with a density change
$\sqrt{4/3} \rightarrow 2\sqrt{4/3}$. To introduce a little chaos into
the initial conditions random initial velocities, corresponding to a
temperature $10^{-10}$ were chosen.
Because the initial pressure is zero the conservation relations are as
follows:
$$
\rho v = \sqrt{4/3} \times 1.930 = 2.229 \ ;
$$
$$
\ P_{xx} + \rho v^2 = \sqrt{4/3} \times 1.930^2 = 4.301 \ ;
$$
$$
\rho v[e + (P_{xx}/\rho) + \rho v^2/2)] + Q_x =
\sqrt{4/3} \times 1.930^3/2 = 4.151 \ .
$$

Although the reversibility of the dynamics cannot be perfect, the shockwave
propagates so rapidly that a visual inspection of the reversed dynamics
shows no discrepancies over thousands of Runge-Kutta timesteps.  To
assess the mechanical instability of the shock compression process we
explore the effects of small perturbations to the reversible dynamics in
the following Sections. We begin by illustrating phase-space
instability\cite{b26,b58} for a simpler problem, the harmonic chain.

\subsection{Linear Growth Rates for a Harmonic Chain}

Even the one-dimensional harmonic chain, though not chaotic,  exhibits
linear phase-volume growth in
certain phase-space directions.  Consider the equations of motion for a
periodic chain incorporating an arbitrary scalefactor $s^{+2}$:
$$
\{ \ \dot q = ps^{+2} \ ; \ \dot p = (q_+ - 2q + q_-)s^{-2} \ \} \ ;
$$
the subscripts indicate nearest-neighbor particles to the left and right.
The motion equations for a $2N$-dimensional perturbation vector
$\delta = (\delta q,\delta p)$
follow by differentiation:
$$
\{ \ \delta \dot q = \delta ps^{+2} \ ; \ 
\delta \dot p =(\delta q_+ - 2 \delta q +  \delta q_-)s^{-2} \ \} \ .
$$
If we choose the length of the perturbation vector equal to unity,
the logarithmic growth rate, $\Lambda = (d\ln \delta/dt)_{q,p}$,  is
a sum of the individual particle
contributions:
$$
\Lambda (\delta ) = \sum [ \delta q\delta p(s^{+2} - 2s^{-2}) +
\delta p(\delta q_+  + \delta q_-)s^{-2} ] \ .
$$
For a {\em large} scale factor $s^{+2}$ it is evident that choosing
equal components of the vector provides the maximum growth rate,
$$
\{ \delta q = \delta p = \sqrt{1/2N} \} \ \rightarrow \
\Lambda _{\rm max} = 2^{-1}s^{+2} \ .
$$
For $s^2$ small, rather than large, alternating signs give the largest growth
rate, with
$$
\{ \ +\delta q_{\rm even} = +\delta p_{\rm odd} = -\delta q_{\rm odd} = -\delta
p_{\rm even} \ \} \ ,
$$
the growth rate is 
$$
\Lambda _{\rm max} = 2^{+1}s^{-2} - 2^{-1}s^{+2}  \ .
$$
The growth rate is $2^{-1/2}$ at the transition between the two regions,
where $s^2 \stackrel{<}{>} 2^{1/2}$.

These same growth-rate results can be found numerically by applying
``singular value decomposition'' to the dynamical matrix $D$\cite{b26,b58}.
This analysis details the deformation of an infinitesimal phase-space
hypersphere for a short time $dt$.  During this time the hypersphere
has its components $\delta q,\delta p$ changed by the equations of motion:
$$
\delta \stackrel{dt}{\longrightarrow} (I + Ddt)\cdot\delta \ ,
$$
so that the growth and decay rates can be found from the diagonal elements
of the singular value decomposition
$$
I + Ddt = U\cdot W\cdot V^t \rightarrow \{ \Lambda = (1/dt)\ln W \} \ .
$$
Numerical evaluation gives the complete spectrum of the growth and decay
rates.  The maximum matches the  analytic results given above.  Although
locally the growth rates $\{ \Lambda(r,t) \}$ are nonzero, the harmonic chain
is not at all chaotic and the long-time-averaged Lyapunov exponents
$\{ \lambda = \langle \lambda (r,t) \rangle \}$, all vanish.  Let us now
apply the concepts of phase-space growth rates $\{ \Lambda \}$ and
the Lyapunov exponents $\{ \lambda \}$ to the shockwave problem.

\subsection{Linear Instability in Many Body Systems, $\Lambda $ for Shockwaves}

\begin{figure}
\vspace{1 cm}
\includegraphics[height=15cm,width=10cm,angle=-90]{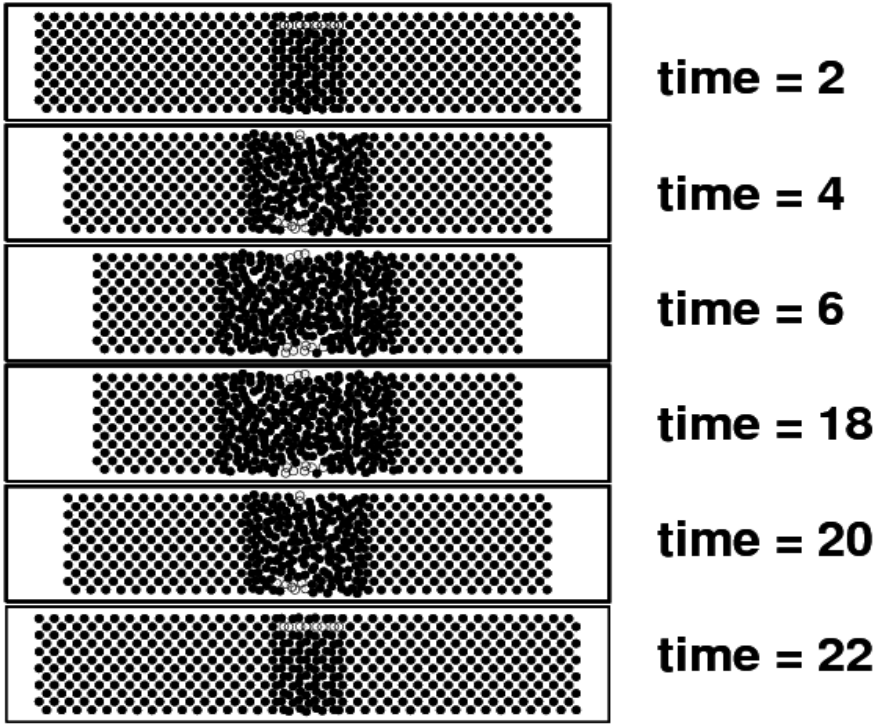}
\caption{
Phase-space Growth Rates $\{ \Lambda \}$ during the collision of two
240-particle blocks of length 20.  The collision leads to twofold compression
of the original cold material at a time of order $20/1.93 \simeq 10.4$.
At time = 12 the velocities were reversed, so that the configurations
at times 2, 4, and 6 correspond closely to those at 22, 20, and 18
respectively.  Those particles making above average contributions to
the largest phase-space growth direction are indicated with open circles.
The fourth-order Runge-Kutta timestep is $dt = 0.002$.
}
\end{figure}

The time reversibility of the Hamiltonian equations of motion guarantees
that any stationary situation shows both a long-time-averaged and a local
symmetry between the forward and reversed directions of time.  In such a
case the $N$ nonzero time-averaged Lyapunov exponents as well as the local
growth rates, obey the relations
$$
\{ \ \lambda_{N+1-k} + \lambda_{k} \} = 0 \ ; \
\{ \ \Lambda_{N+1-k} + \Lambda_{k} \} = 0 \ .
$$
The instantaneous Lyapunov exponents $\{ \lambda (t) \} $ depend on the
dynamical history, while the instantaneous diagonalized phase-space growth
rates, which we indicate with $\Lambda(t)$ rather than $\lambda(t)$, do not.

The rates $\{ \Lambda(t) \}$ for different directions in phase space can be
calculated efficiently from the dynamical matrix $D$, by using singular value
decomposition, just as we did for the harmonic chain:
$$
D = 
\left(
\begin{array}{cc}
\partial \dot q/\partial q &\ \partial \dot q/\partial p \\
\partial \dot p/\partial q &\ \partial \dot p/\partial p                      
\end{array}
\right)
=
\left(
\begin{array}{cc}
0 & \ 1/m \\
 \partial F/\partial q & 0                
\end{array}
\right)
\ .
$$

Here we analyze a 480-particle shockwave problem, the collision of two blocks
with $x$ velocity components $\pm 0.965$.  Figure 20 shows those particles
making above-average contributions to the maximum phase-space growth rate
at times 2, 4, and 6. At time 12 the particle velocities are all reversed,
so that the configurations at times 22, 20 and 18 closely match those at
times 2, 4, and 6. Generally there is six-figure agreement between the
coordinates going forward in time and those in the reversed trajectory at
corresponding times.  Note this symmetry in Figure 20, where the most
sensitive particles going forward and backward are exactly the same at
corresponding times.  The forward-backward agreement could be made perfect
by following Levesque and Verlet's suggestion\cite{b59} to use {\em integer}
arithmetic in evaluating a time-reversible (even {\em bit-reversible}!)
algorithm such as
$$
{\rm Int}[q_{t+dt} - 2q_{t} + q_{t-dt}] = {\rm Int}[F_tdt^2/m] \ {\rm or}
$$
$$
{\rm Int}[q_{t+2dt} - q_{t+dt} - q_{t-dt} + q_{t-2dt}] =
{\rm Int}[(dt^2/4m)(5F_{t+dt} + 2F_{t} + 5F_{t-dt})] \ .
$$

Evidently, as would be expected, from their definition, the point-function
growth rates $\{ \Lambda (r(t)) \}$ can  show no ``arrow of time''
distinguishing the backward trajectory from the forward one.  We turn next
to the Lyapunov exponents, which {\em can} and {\em do} show such an arrow.

\subsection{Lyapunov Spectrum in a Strong Shockwave}

Most manybody dynamics is Lyapunov unstable, in the sense that the length
of the phase-space vector joining two nearby trajectories has a tendency to
grow at a (time-dependent) rate $\lambda _1(t)$ (with the time-averaged result
$\lambda _1 \equiv \langle \lambda _1(t) \rangle > 0 $).  Likewise, the area
of a moving phase-space triangle, with its vertices at three nearby
trajectories, grows at $\lambda _1(t) + \lambda _2(t)$, with a time-averaged
rate $\lambda _1 + \lambda _2$.  The volume of a tetrahedron defined by four
trajectories grows as $\lambda _1(t) + \lambda _2(t) + \lambda_3(t)$, and so 
on.  By changing the scale factor linking coordinates to momenta -- the
$s^{+2}$ of the last Section -- these exponents can be determined separately
in either coordinate or momentum space.

Posch and Hoover, and independently Goldhirsch, Sulem, and Orszag, 
discovered a thought-provoking representation of local Lyapunov
exponents\cite{b60,b61}.  If an array of Lagrange multipliers is chosen to
propagate a comoving corotating orthonormal set of basis vectors centered on
a phase space trajectory, the diagonal elements express {\em local} growth
and decay rates.  These are typically quite different (and unrelated) in the
forward and backward directions of time. 

Let us apply the Lyapunov spectrum\cite{b59,b60,b61} to the phase-space
instability of a strong shockwave.  Because the Lyapunov exponents,
$\{ \lambda(t) \}$, are evaluated so as to reflect only the past, times
less than $t$,
we expect to find that the Lyapunov vector corresponding to maximum growth
soon becomes localized near the shock front. Starting out with randomly
oriented vectors the time required for this localization is about 1/2.
The time-linked disparities between the forward and backward motions
suggest that the
Lyapunov exponents {\em can} provide an ``Arrow of Time'' because the
stability properties forward in time differ from those in the backward
(reversed) direction of time\cite{b62}.

 Figure 21 shows the particles making above-average contributions to the
largest of the local Lyapunov exponents, $\lambda_1(t)$.  There are many
more of these particles than the few which contribute to the largest of
the phase-space growth rates, $\Lambda_1(t)$.  The shockwave simulation
was run forward in time for 6000 timesteps, after which the velocities
were reversed.  The phase-space offset vectors, chosen randomly at time
0 and again at time 12, became
localized near the shockfront at a time of order 0.5.  The particles to
which the motion is most sensitive, as described by the Lyapunov exponent
$\lambda _1(t)$ are more localized in space in the forward direction of
time than in the backward direction.  Evidently the Lyapunov vectors are
more useful than the vectors corresponding to local growth rates in
describing the irreversibility of Hamiltonian systems.

\begin{figure}
\vspace{1 cm}
\includegraphics[height=15cm,width=10cm,angle=-90]{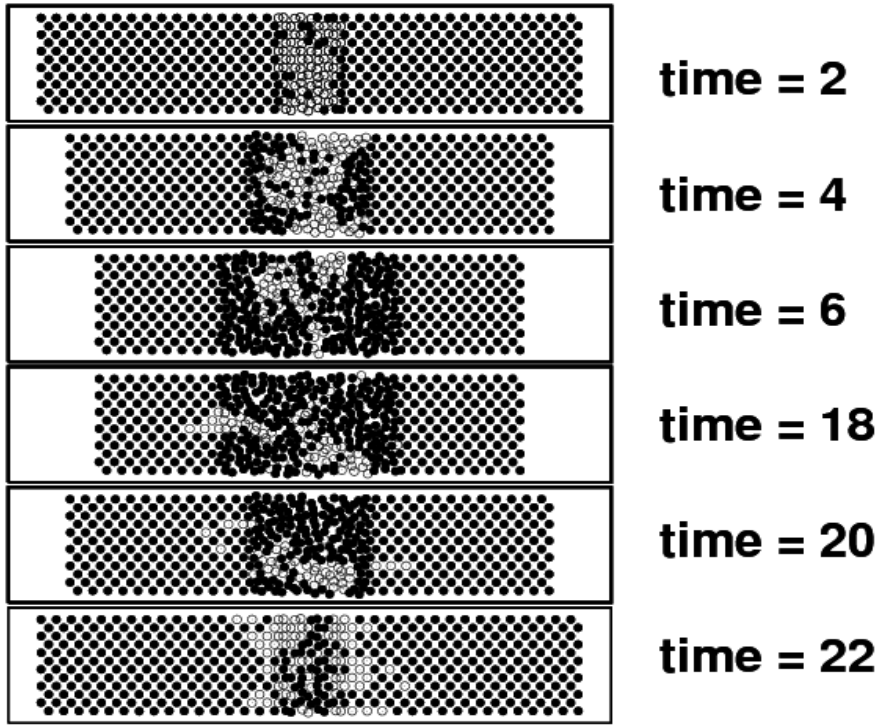}
\caption{
Phase-space Growth Rates $\{ \lambda \}$ during the collision of two
similar blocks which lead to the twofold compression of the original
cold material.   At time = 12 the velocities were
reversed, so that the configurations at times 2 and 4 correspond to those
at 22 and 20, respectively.  The particles making above average contributions
to the largest Lyapunov exponent are indicated with open circles.
}
\end{figure}

\section{Conclusion}

Particle dynamics, both NEMD and SPAM, provides a flexible approach to the
simulation, representation, and analysis of nonequilibrium problems.  The
two particle methods are closely
related, making it possible to infer constitutive relations directly from
atomistic simulations.  These useful tools provide opportunities for
steady progress in understanding far-from-equilibrium states.  It is our
hope that these tools will become widely adopted.

\section{Acknowledgment}

We thank Vitaly Kuzkin for encouraging our work on this review.

\end{document}